\documentclass[11pt,twoside]{article}
\usepackage{asp2004}
\usepackage{psfig}
\usepackage{epsf}
\usepackage{graphics}
\usepackage{graphicx}
\usepackage{lscape}
\def\edcomment#1{\iffalse\marginpar{\raggedright\sl#1\/}\else\relax\fi}
\reversemarginpar

\begin{document}
\title{Intermediate-band Photometry of Type Ia Supernovae}
\author{X. F. Wang, X. Zhou, T. M. Zhang}
\affil{National Astronomical Observatories of China, Chinese
Academy of Sciences, Beijing 100012, PR China}
\author{Z. W. Li}
\affil{Department of Astronomy, Beijing Normal University, Beijing
100875, P R China}

\begin{abstract}
We present optical light curves of five Type Ia supernovae
(2002er, 2002fk, 2003cg, 2003du, 2003fk). The photometric
observations were performed in a set of intermediate-band filters.
SNe 2002er, 2003du appear to be normal SN Ia events with similar
light curve shapes, while SN 2003kf shows the behavior of a
brighter SN Ia with slower decline rate after maximum. The light
curves of SN 2003cg is unusual; they show a fast rise and dramatic
decline near maximum and do not display secondary peak at longer
wavelengths during 15-30 days after maximum light. This suggests
that SN 2003cg is likely to be an intrinsically subluminous,
91bg-like SN Ia. Exploration of SN Ia feature lines through
intermediate-band photometry is briefly discussed.
\end{abstract}
\thispagestyle{plain}

\section{Introduction}
Supernova (SN) light curves are one of the major sources of
information about the nature of these events. They infer
structures of the stellar progenitors and reflect the underlying
energy sources created in the explosion. Direct comparison of peak
luminosities and multiband light curves with the predictions of
theoretical models lead to better understanding of the physics of
supernova explosion. Light curves of thermonuclear supernovae (SNe
Ia) are now the primary tool for the study of precise cosmology
\citep{lei01}.

We have been engaged in systematic optical monitoring of bright
SNe in nearby galaxies (z$<$0.01). The photometric system we use
is part of 15-filter intermediate-band system designed to cover
the wavelength range 3300 $\AA$- 1$\mu$m, which avoids known and
variable sky emission. A description of this unique photometric
system, also dubbed the BATC system, are detailed elsewhere
\citep{fan96, yan99}. By using the intermediate-band filters we
may learn more about SN light curves. In particular, we could
explore the evolution of some spectrum lines during SN explosion
through spectrophotometry. Here we report our initial results of
the intermediate-band SN photometry program.

\section{Intermediate-band Light Curves}
In Figure 1, we present the intermediate-band light curves of five
SNe Ia. They are SN 2002er in UGC 10743, SN 2002fk  in NGC 1309,
SN 2003cg in NGC 3169, SN 2003du in UGC 9391, and 2003kf in
MCG-02-16-002. Observations of these SNe were made near maximum
light in several BATC filters (e.g. $\emph{d}$th, $\emph{h}$th,
$\emph{i}$th, $\emph{k}$th, and $\emph{n}$th filters that centered
on 4550$\AA$, 6075$\AA$, 6660$\AA$, 7490$\AA$ and 8480$\AA$
respectively). As can be seen from the figure, the light curves of
SNe Ia in d and h filters resembles those in B and V except for a
decrease of the decline after maximum. The secondary peak feature
was observed in BATC's redder light curves. The most pronounced
feature presents in k band, where no prominent spectrum line
emerges during the second peak period. It is therefore that the
occurrence of the secondary peak is not due to the spectral
feature but it is likely due to the time evolution of line
opacities \citep{whe98}

\begin{figure}[h]
\vskip -0.7cm
\includegraphics[width=13.1cm,angle=0]{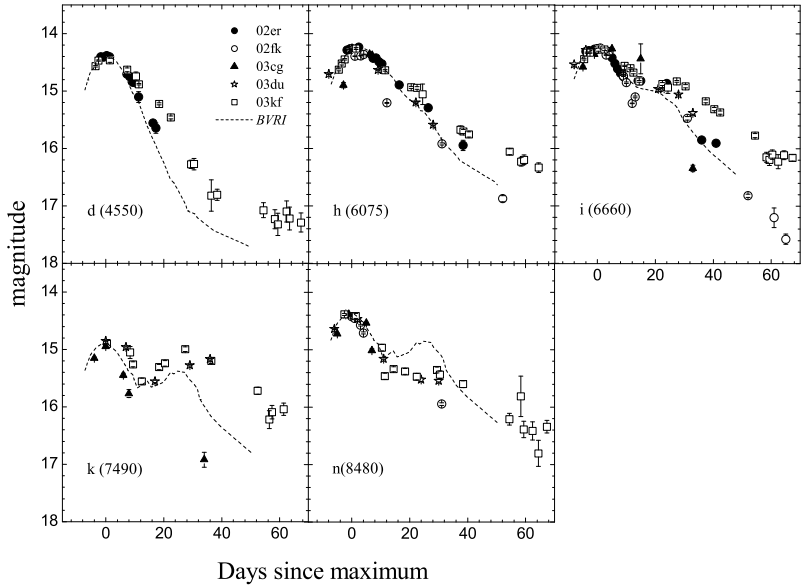}
\vskip -0.6cm \caption{Intermediate-band light curves of SNe Ia.
Different symbols refer to different SNe. The dashed lines
represent the broad band BVRI light curves of SN 2002er (Pignata
et al. 2004). For comparison, the light curves of other four SNe
Ia and BVRI for SN 2002er are shifted to get the best match to the
well-sampled SN 2003kf between $-$7.0 and +7.0 days.} \vskip
-0.2cm \label{fig1}
\end{figure}

Nevertheless, the diversity of the light curves in several of
BATC's passbands is readily apparent. The differences are most
pronounced in i band (and possibly in k band), where the
differences in magnitude decline after maximum could reach
$\sim$0.5 mag at epoch of 30 days and reach $\sim$1.0 mag at 60
days even without including SN 2003cg. We considered SN 2003cg as
an intrinsically subluminous SN Ia for two reasons. First, the
shoulder or secondary peak appears to be missing for SN 2003cg in
i and k light curves. This feature is absent for SN 1991bg-like SN
Ia events such as SNe 1997cn \citep{tur98}, 1999by \citep{gar01}.
Second, the light curves of SN 2003cg near maximum light are much
narrower than normal SNe Ia (see Fig 1). On the other hand, SN
2003kf seems to be an luminous SN Ia. The light curves of SN
2003kf in d band declined by 0.68$\pm$0.02 mag in the first 15
days after maximum, while the counter value of SN 2002er is
1.03$\pm$0.03 mag and its $\Delta$$m_{15}$ is reported to be
1.33$\pm$0.04 \citep{pign04}.

\section{Evolution of Feature Lines}
One merit of using BATC's filters is that they may properly cover
the broad spectral lines of SNe at proper redshift. For instance,
the h filter may cover Si II $\lambda$6355 (blueshifted to
6150$\AA$) and the n filter may cover CaII IR triplets for nearby
SNe Ia at $z<0.01$. Thus we can gain some knowledge about the
evolution of some feature lines of supernovae from the
corresponding intermediate-band light curves.


\begin{figure}[h]
\vskip -0.7cm
\includegraphics[width=13.1cm,angle=0]{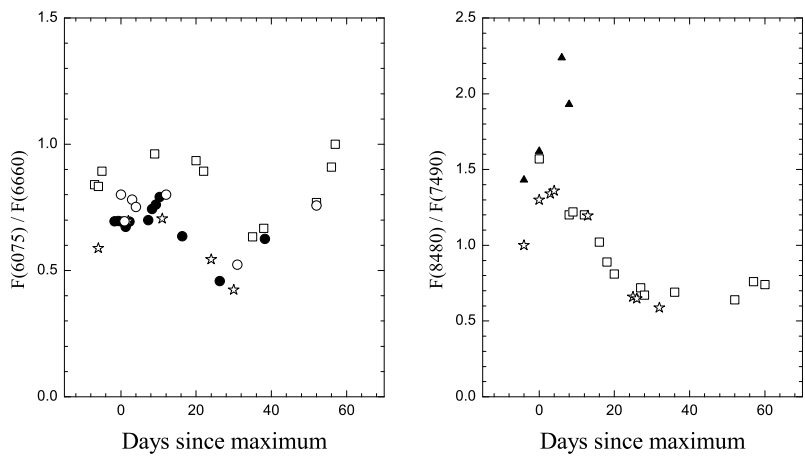}
\vskip -2.2cm \caption{Flux ratios of four BATC's
intermediate-band filters. The symbols in this figure are the same
as in Fig 1.}
 \vskip -0.2cm
\label{fig1}
\end{figure}

Figure 2 shows the flux ratio of h band vs. i band and n band vs.
k band. F(6075)/F(6660) may indicate the variation of
SiII$\lambda$6355 while F(8480)/F(7490) may reflect the evolution
of CaII IR triplet during the explosion.

The observations discussed here are part of our intermediate-band
photometric program on bright SNe. The intermediate-band
photometric data of a total 20 SN sample are being analyzed for
further publication (Wang et al. 2004 in preparation).

\acknowledgments Financial support for this work has been provided
by the National Science Foundation of China (NSFC grant 10303002;
10173003) and National Key Basic Research Science Foundation
(NKBRSF TG199075402).

\end{document}